\documentclass[11pt]{article}
\newcommand{\be}{\begin{equation}}
\newcommand{\ee}{\end{equation}}
\newcommand{\bq}{\begin{quote}}
\newcommand{\eq}{\end{quote}}
\newcommand{\bi}{\begin{itemize}}
\newcommand{\ei}{\end{itemize}}
\newcommand{\ben}{\begin{enumerate}}
\newcommand{\een}{\end{enumerate}}

\newcommand{\smiley}{\begin{tabular}{l}\raisebox{-2pt}{$\odot$}%
   \\\raisebox{2pt}{$\odot$}\end{tabular}$\!\!\!\rightharpoonup${\large)}}

\begin{document}
\title{Comment on ``Quantum Physics from A to Z''}
\author{Ulrich Mohrhoff\\
Sri Aurobindo International Centre of Education\\
Pondicherry 605002, India\\
\textsf{ujm@auromail.net}}
\date{}
\maketitle
\begin{abstract}
\noindent This is a comment on a collection of statements gathered on the occasion of the \textit{Quantum Physics of Nature} meeting in Vienna.
\end{abstract}
\vspace{10pt}\noindent In a recent paper~\cite{a-z}, twenty-seven authors presented their personal views on Anton Zeilinger's philosophy of physics with a view to stimulating a discussion. Here is my humble contribution toward this worthy objective.
\bi\item There is no quantum information. There is only a quantum way of handling information. (Attributed to A.~Z.)
\ei We have two kinds of information about the quantum world, (i)~information of nomological character---the quantal correlation laws---and (ii)~information about matters of fact---the outcomes of performed measurements. Together they permit us to assign probabilities to the possible outcomes of (as yet) unperformed measurements. From here on, there is no need to ever again use the word ``information.''

In the good old days of classical physics, all there was, was deterministic correlations between measurement outcomes. Because they were deterministic, it was OK to imagine causal links, but the idea that we had any understanding of the nature of the causal links (beyond the correlations) was a delusion. The transmogrification of mathematical symbols into physical processes or states of affairs was never more than a sleight of hand. All we now have is probabilistic correlations between measurement outcomes. With probabilistic correlations, the transmogrification of mathematical symbols into physical processes or states of affairs no longer works.%
\footnote{So it's harder for a teacher to avoid pesky student questions, albeit not for want of trying. In the Prologue to their beautiful book \textit{The Quantum Challenge}, Greenstein and Zajonc~\cite{GZ} observe that ``in every textbook we know, quantum mechanics has been largely sanitized of these beautiful enticements and their implications.'' It makes one wonder\dots}
\bi\item Photons are clicks in photon detectors. (Attributed to A.~Z.)
\ei Photons are \textit{correlations between} clicks in counters. 
\bi\item Descriptions of experiments with photons should not make use of the term vacuum. (Attributed to A.~Z.)
\ei That term should be eliminated from the dictionary of quantum physics. Positions are not different from other properties: if they are not possessed then they do not exist. There is no empty space because there are no unpossessed positions. Where there is nothing, there is no \textit{there}. (More carefully worded: positions only exist as spatial relations, and these do not exist without material relata.)
\bi\item Objective randomness is probably the most important element of quantum physics. (Attributed to A.~Z.)
\ei And understanding objective randomness is certainly the most urgent task for quantum philosophers. Unfortunately, most of their time is wasted with pseudoproblems. As a philosopher of science once put it to me, to solve the PLOP%
\footnote{The piddling laboratory operations problem: ``To restrict quantum mechanics to be exclusively about piddling laboratory operations is to betray the great enterprise.'' (John Bell~\cite{Bell1990})}
``means to design an interpretation in which measurement processes are not different in principle from ordinary physical interactions.'' Immanuel Kant admitted to having been woken from his ``dogmatic slumbers'' by David Hume. This led Bertrand Russell to quip that Kant quickly invented a soporific that allowed him (Kant) to go back to sleep. What said philosopher of science expects from us is another soporific. What we really need to understand is \textit{why} the fundamental theoretical framework of physics is an algorithm for assigning probabilities to possible measurement outcomes (on the basis of actual measurement outcomes), and \textit{why} measurements do play a special role in our fundamental theory of matter.
\bi\item Quantum mechanics applies to single quantum systems. (Attributed to A.~Z.)
\ei
Quantum mechanics applies to \textit{correlations between (any number of) measurement devices}.
\bi\item The border between classical and quantum phenomena is just a question of money. (Attributed to A.~Z.)
\ei A question of brawn rather than a question of brains? I have my doubts.
\bi\item The speed of the collapse is bull \dots
\ei Although this item was censored (commented out in the source file) I couldn't agree more: with or without speed, collapse is [\textit{bleep}].
\bi\item Objective (absolute) randomness is hardly fully verifiable\dots the a priori exclusion of any reason whatsoever cannot be falsified/verified. (Markus Arndt)
\ei 
True, absence of evidence is not the same as evidence of absence. But we have more than absence of evidence. Even Bohmians agree that their surreal particle trajectories are unobservable. The absence of evidence is \textit{nomological}. Why are Bohmian trajectories unobservable? The simplest answer is: because they don't exist!

But the real issue is not objective randomness; it is objective \textit{fuzziness}.%
\footnote{``Fuzziness'' is the actual meaning of Heisenberg's original term, \textit{Unsch\"arfe}. The stability of atoms rests on the objective fuzziness of their internal relative positions and momenta, not on our ``uncertainty'' about the exact values of these observables.}
In the laboratory, the objective fuzziness of an observable (or its value) evinces itself as a probability distribution (with nonzero dispersion) over alternative outcomes in a series of measurements performed on an ensemble of ``identically prepared'' systems.%
\footnote{Strictly speaking, what evinces itself is the fuzzy value that \textit{would} have obtained if \textit{no} measurement had been made.}

And beyond that, the real issue is to make sense of fuzzy values. To my mind, there is a straightforward way of doing this~\cite{Mohrhoff04,Mohrhoff05,Mohrhoff06}: Whenever quantum mechanics instructs us to add amplitudes rather than probabilities, the distinctions we make between the corresponding alternatives/possibilities are distinctions that Nature does not make. They correspond to nothing in the real world. They exist solely in our heads. The reason why an electron (or buckyball, for that matter) can go simultaneously through two (or many) slits \textit{without} being split into parts that go through different slits%
\footnote{We don't picture parts of a C$_{60}$ molecule as getting separated by many times the period of the diffraction grating (100\thinspace nm) and then reassemble into a ball less than a nanometer across.}
is that the distinctions \textit{we} make between the regions defined by the slits (or between the alternative paths) correspond to nothing in the real world---as far as that electron or buckyball is concerned. The reality of spatial distinctions is \textit{relative}: the distinction we make between two regions or paths may be real for one system at one time and not real for another system at the same time or for the same system at another time. From this it follows that \textit{space must not be thought of as consisting of (intrinsically distinct) parts.}

And beyond that, the real issue is: What are the conditions under which those distinctions are real (for a given system at a given time)? In other words, when does a proposition like ``the electron went through the left slit'' have a truth value? As far as unadulterated, standard quantum mechanics is concerned---no surreal particle trajectories {\it\`a la\/} Bohm~\cite{Bohm52}, no nonlinear modifications of the Schr\"odinger equation {\it\`a la\/} Ghirardi, Rimini, and Weber~\cite{GRW86} or Pearle~\cite{Pearle89}, no extraneous axioms like the eigenstate-eigenvalue link~\cite{vF91} or the modal semantical rule~\cite{Dieks94}---the only condition available is to be measured. No property is a possessed property unless its possession can be inferred from some actual event or state of affairs---a ``measurement.''

The implied \textit{supervenience} of the microscopic on the macroscopic contradicts one of our most incorrigible ``intuitions,'' according to which the properties of macroscopic objects exist by virtue of the properties of microscopic objects: the macroworld is what it is because its microscopic constituents are what they are. The ontological implications of the quantal probability algorithm are incompatible with this bottom-up philosophy. The statistics of indistinguishable particles does not permit us to build reality out of a multitude of intrinsically distinct, transtemporally identical material parts, and the relativity of the reality of spatial distinctions does not permit us to build reality out of (or on) intrinsically distinct spatial parts. The ``foundation'' \textit{is} the macroworld.

Once we get used to \textit{not} thinking of physical space as consisting of intrinsically distinct parts or, God help us, points, it is easy to see that we need a macroscopic detector not only to indicate the presence of something somewhere but also to realize (make real) the property of being inside the detector's sensitive region---to make a position available for attribution.

\bi\item I think that no object can be defined without reference to its external world. (Markus Arndt)
\ei The way I see it, no microscopic object can be defined without reference to the macroworld, and no macroscopic object can be defined without reference to the \textit{rest} of the macroworld. Macroscopic objects, properly 
defined~\cite{Mohrhoff04,Mohrhoff05}, localize each other (relative to each other) so sharply that their relative positions are only counterfactually fuzzy. They are fuzzy only relative to an imaginary background that is more differentiated spacewise than is the real world.
\bi\item The fascination of quantum physics lies in the fact that it is not yet formulated coherently by a set of simple foundational principles. (Markus Aspelmeyer)
\ei It probably cannot be so formulated, for the same reason that the quantum world cannot be built from the bottom up.
\bi\item There is no border between classical and quantum phenomena -- you just have to look closer. (Reinhold Bertlmann)
\ei There is no border, but there is a limit. In a world of ``more or less fuzzy,'' the \textit{least} fuzzy is (onto)logically different from the rest.
\bi\item We don't know why events happen. (A.~Z. as quoted by Chris Fuchs)
\ei Quantum physics concerns correlations between property-indicating events or states of affairs. It does not account for the occurrence or existence of the correlata, anymore than classical physics explains why there is anything, rather than nothing at all. On the other hand, as ingredients of a theory that presupposes correlata, the quantal correlation laws are trivially consistent with their occurrence or existence.
\bi\item We have to understand therefore what it means to collect information about something which is not as much structured as we think.\hfill\break
(A.~Z. as quoted by Chris Fuchs)
\ei That the world is not as much structured as we think seems to me to be a crucial point. If we go on dividing a material object, its so-called ``constituents'' lose their individuality, and if we conceptually partition the world into sufficiently small but still finite regions, we reach a point where the distinctions we make between regions of space no longer correspond to anything in the physical 
world~\cite{Mohrhoff04,Mohrhoff05,Mohrhoff06}. Our spatial and substantial distinctions are warranted by property-indicating events, and these do not license an absolute and unlimited objectification of spatial or substantial distinctions.
\bi\item I believe that there is \textit{no} classical world. There is \textit{only} a quantum world.\hfill\break
(Dan Greenberger)
\ei Yes and No! In a world that is not completely differentiated spacewise (nor timewise), the least fuzzy is only \textit{counter}factually fuzzy and therefore \textit{factually} sharp; it conforms to \textit{both} quantum mechanics \textit{and} classical laws. Differently put, in the case of macroscopic objects, the quantal probability algorithm (represented by a ray in a vector space) degenerates into the classical probability algorithm (represented by a point in a phase space). Because this assigns only trivial probabilities (1~or~0), it can be re-interpreted as an evolving state of affairs. Because quantal probabilities are generally nontrivial, the corresponding re-interpretation of the quantal algorithm is impossible.

\bi\item Quantum mechanics is magic! (Dan Greenberger)
\ei Absolutely! There is no way of knowing ``how Nature does it''---for the simple reason that the implications of a fundamental theory can never be explained by a ``more fundamental'' theory. (If there were such a theory, the ``less fundamental'' theory would not be fundamental at all.) But why should this come as a surprise? Because classical physics seemed to permit the transmogrification of mathematical symbols into physical processes or states of affairs? Let's get real!
\bi\item But what about the energy flowing from the source to the detector?\hfill\break
(Thomas Jennewein)
\ei We have a probability algorithm that correlates an event involving the source with an event involving the detector. This probability algorithm features an action. This action is ``local'' (i.e., a spacetime integral over a Lagrange density) in order to ensure (for every inertial frame) that effects are later than their causes.%
\footnote{I am talking about controllable effects, not spooky passion at a distance.}
Noether's theorem guarantees a conservation law for every continuous symmetry of the Lagrange density, in particular the conservation of energy-momentum, which is a trivial consequence of a sensible choice of coordinates (namely, inertial coordinates). Does it follows from this that energy-momentum is some kind of physical fluid sloshing about in spacetime? Don't forget that we are talking \textit{exclusively} about the mathematical features of a probability algorithm!
\bi\item Would Anton agree that electrons are clicks in electron counters? Are fullerenes clicks in fullerene counters? Is Anton a click in an Anton counter?\hfill\break
(N. David Mermin)
\ei I don't think a click in an Anton counter would agree that electrons are clicks in electron counters. But Anton might.%
\footnote{If I am correctly informed, Anton is a likable intelligent conscious living being. I refuse to join in the pretense that such beings form part of the subject matter of physics.}
At any rate, I do. However, conservation laws make a difference:

\textit{Case~1}: A photosource mysteriously loses a certain (more or less fuzzy) energy-momentum, a photodetector mysteriously gains the same (after a suitable interval depending invariantly on the distance between the two events). Due to our penchant for confabulation, we speak of the emission, propagation, and absorption of a photon. Because there is no photon before the so-called emission or after the so-called absorption (at any rate, not the \textit{same} photon), the chief merit of this story is that it gives rise to entertaining puzzles about the propagation of photons.

\textit{Case~2}: As long as (or to the extent that) there is a conservation law for electrons, an electron exists both before and after its detection. However, it only exists because its existence is indicated by clicks in electron counters, and the properties it has are precisely those whose possession is indicated by the goings-on in the macroworld. The same holds for fullerenes, as long as there is an effective conservation law for fullerenes. Under conditions in which fullerenes break up or otherwise react chemically, the story gets more complicated, but nothing alters the fact that what happens is what can (in principle) be inferred from the goings-on in the macroworld.

\textit{Case~3}: Apart from energy-momentum, only charges are conserved. In this (fully relativistic) case, quantum mechanics does not deal with systems of this or that kind, for the permanent labeling of a system as being of this or that kind requires an effective conservation law, e.g., a conservation law for Antons, which exists only under conditions favorable to the survival of Antons.
\bi\item Chris Fuchs has taught me to beware of conjoining ``objective'' to
``probability''. (N. David Mermin)
\ei Chris Fuchs has a way of defining ``probability'' that makes it inherently subjective. According to Fuchs and Peres%
\footnote{\textit{De mortuis nil nisi bene}: With statements like the following, Asher Peres has earned my deep and abiding respect: ``\dots there is no interpolating wave function giving the `state of the system' between measurements.''~\cite{Peres}}
\cite{FuPer}, ``probability theory is simply the quantitative formulation of how to make rational decisions in the face of uncertainty.'' The quantal correlation laws do not depend on the existence of rational decision makers. Quantal probabilities are objective probabilities (not to be confused with relative frequencies) because they are irreducible ingredients in our fundamental theory of the objective world. They are macroscopic manifestations of the objective fuzziness that fluffs out matter.

However, in at least one important respect I agree with Chris Fuchs and N. David Mermin: there are not \textit{absolute} probabilities. Kolmogorov~\cite{Kol50} defines conditional probabilities in terms of absolute, a~priori probabilities. As far as quantum mechanics is concerned, this puts the cart in front of the horse. It bamboozles one into thinking that wave functions and state vectors (i)~define absolute probabilities and (ii)~exist in an anterior relationship to propagators (which define conditional probabilities). As emphasized by Primas~\cite{Primas03}, quantal probabilities are \textit{conditional} probabilities.%
\footnote{A more natural formulation of probability theory from the quantal point of view is therefore that due to R\'enyi~\cite{Renyi55}, which is based on conditional probabilities. R\'enyi~\cite{Renyi70} has shown that every result of Kolmogorov's probability theory can be translated into a theory based entirely on conditional probabilities.}

Consider, for instance, a stationary state of atomic hydrogen. As every student of quantum mechanics can confirm, one readily lapses into thinking that this is an objective state defining absolute probabilities, whereas in reality it is an algorithm that serves to compute conditional probabilities, e.g., the probability of finding the electron in a certain region of the imagined space of exact positions relative to the proton, \textit{given that} the atom was previously subjected to measurements of three observables: its energy, its total angular momentum, and one component of its angular momentum.

The misconception that wave functions define absolute probabilities gains support from a hangover from classical physics, the \textit{evolutionary paradigm} according to which physics can be neatly divided into kinematics (concerning the description of a physical system at any one time) and dynamics (concerning the evolution of a physical system from earlier to later times). A world that is not completely differentiated spacewise is also not completely differentiated timewise~\cite{Mohrhoff04,Mohrhoff05}. And in a world that is not completely differentiated timewise, evolution is ill-defined. To the question of how many modes of evolution there are, the correct answer is neither two (unitary and state reduction) nor one (only unitary) but \textit{none}.

The evolutionary paradigm conceals a most important aspect of the quantal correlation laws: their time-symmetry. They allow us not only to assign prior probabilities (on the basis of earlier measurement outcomes) but also posterior probabilities (on the basis of later outcomes) and even probabilities that depend on both earlier and later outcomes. The latter are calculated according to the ABL rule~\cite{ABL64,Mohrhoff01} rather than the standard Born rule, and they can be encapsulated in a ``two-state''~\cite{AV91} in place of a wave function. After a spin-0 particle has decayed into two spin-$\frac{1}{2}$ particles, the spins of the latter are ``entangled'' (meaning that outcomes of spin measurements with respect to the same axis are anticorrelated)---and so are the spins of two particles that are yet to fuse into a spin-0 particle!

\bi\item After all, we understand very little about our universe! (Christophe Salomon)
\ei How true!
\bi\item If we want to look beyond, we have to find a completely new way to look at nature. (J\"org Schmiedmayer)
\ei Perhaps the way of looking I've been trying to communicate for some time? \ \smiley
\bi\item Reversibility is just a question of money. (Vlatko Vedral)
\ei Negative. Reversibility requires determinism, which requires a world that is completely differentiated both spacewise and timewise (that is, a world unlike ours).
\bi\item Physics will in the future put less emphasis on equations and mathematics but more on verbal understanding. (A.~Z. as quoted by Peter Zoller)
\ei Obviously. Mathematical equations do not address the problem of making physical sense of mathematical equations.

\end{document}